\documentclass[preprint,aps]{revtex4}
\usepackage{amsmath}
\usepackage{graphicx}
\usepackage{dcolumn}
\usepackage{bm}

\begin{document}

\title{The Intermodulation Coefficient of an
Inhomogeneous Superconductor}

\author{Kwangmoo Kim and David Stroud}
\affiliation{Department of Physics, The Ohio State University, Columbus,
OH 43210, USA}
\date{\today}
\begin{abstract}
The high-T$_c$ cuprate superconductors are now believed to be
intrinsically inhomogeneous.  We develop a theory to describe how
this inhomogeneity affects the intermodulation coefficient of such a
material.  We show that the continuum equations describing
intermodulation in a superconducting layer with spatially varying
properties are formally equivalent to those describing an
inhomogeneous dielectric with a nonzero cubic nonlinearity.  Using
this formal analogy, we calculate the effect of inhomogeneity on the
intermodulation coefficient in a high-T$_c$ material, using several
assumptions about the topology of the layer, and some simple
analytical approximations to treat the nonlinearity. For some
topologies, we find that the intermodulation critical supercurrent
density $J_{IMD}$ is actually {\it enhanced} compared to a
homogeneous medium, thereby possibly leading to more desirable
material properties.  We discuss this result in light of recent
spatial mappings of the superconducting energy gap in BSCCO-2212.

\end{abstract}


\maketitle

\section{\label{sec:level1}Introduction}

Ever since the discovery of high-T$_{c}$
superconductors \cite{bednorz}, many workers have attempted to
develop practical applications for them.  One potential
electronic application is as a microstrip resonator.  Such a device
has been developed by Willemsen {\it et al.} \cite{willemsen},
using a high-T$_c$ cuprate material.
Even though these devices do not work
at very high current densities, they are subject to strong nonlinear
effects which mix together microwaves of different frequencies. 
This mixing, known as intermodulation, 
was studied experimentally and theoretically by Willemsen {\it et al.}
\cite{willemsen}; theoretical models to explain their measurements
were developed by Dahm and Scalapino \cite{dahm1,dahm2}.

In the model of Refs.\ \cite{dahm1} and \cite{dahm2}, the
intermodulation is described in terms of an intermodulation current
density $J_{IMD}$, which depends on both
temperature and the angle between the direction of current flow and
the crystal axes. An equivalent quantity was
also considered by Yip and Sauls \cite{yip} for a $d$-wave
superconductor. Because of low-lying
quasiparticles, they found that $J_{IMD}$ was much smaller than for
a corresponding $s$-wave superconductor and also that it depended on
the angle between the in-plane current density and the ${\bf
k}$-vector of the gap nodes.  Several other workers have also
studied this intermodulation and the
harmonic generation due to nonlinear effects
\cite{xu,stojkovic,mcdonald}.

In all this work, the high-T$_c$ cuprate
superconductor was assumed to be {\it homogeneous} -- that is, 
the CuO$_2$ layer properties were assumed independent of
position within the layer.  However, recent experimental work has 
invalidated this picture. Specifically, in optimally oxygen-doped 
\cite{davis1}, underdoped, and
slightly overdoped \cite{davis2} 
Bi$_{2}$Sr$_{2}$CaCu$_{2}$O$_{8+\delta}$ (usually called
BSCCO-2212), the superconducting energy gap was found to be 
{\it spatially varying}. This result was obtained from scanning tunneling
microscopy/spectroscopy (STM/S) images of the superconducting
layers. The differential tunneling conductance spectra of the sample
were measured and the position-dependent gap $\Delta$ was inferred
from measurements of the energy difference between two coherence
peaks in the spectra above and below the Fermi level. In the
underdoped sample, the gap was found to map into two distinct
regions. One (called the $\alpha$ domains) had a gap $\Delta <
50\:\mathrm{meV}$; the other (denoted the $\beta$ domains) had
$\Delta > 50\:\mathrm{meV}$. The $\alpha$-domains were identified as
superconducting, because they showed coherence peaks in the
tunneling spectra.
The $\beta$-regions were found to be non-superconducting, and were
identified as a pseudogap phase \cite{davis2}, with a large gap. It
was concluded that the spatially varying superconducting energy gaps
do not arise from impurities, but are instead intrinsic properties
of the material.  Thus, underdoped BSCCO-2212 is an intrinsically
granular superconductor.

In this paper, we consider how this inhomogeneity affects the
intermodulation in a high-T$_c$ superconductor, such as BSCCO-2212.
As implied above, this intermodulation coefficient is actually a
nonlinear transport coefficient.  In fact, this coefficient
describes the current-dependence of the superfluid density in the
superconductor. This current-dependence is particularly strong in
the $d$-wave high-T$_c$ materials, because 
quasiparticles are excited even at very low applied currents.
We will show that $J_{IMD}$ is formally analogous
to another coefficient, well known in the study
of nonlinear dielectrics.  Using this connection, we will
demonstrate that $J_{IMD}$ is very sensitive
to the detailed geometry of the superconducting inhomogeneity.

We will consider only two-dimensional (2D) systems,
and only very low frequencies.
This regime is appropriate to the cuprate superconductors, where
superconductivity is believed to occur within CuO$_{2}$ planes. 
Our low-frequency approach should be applicable so long as the length scale of the
inhomogeneity is much smaller than the wavelength of the applied
microwave fields, a condition easily satisfied at microwave frequencies.

The remainder of this paper is organized as follows. In Section
\ref{sec:level2}, we present the formalism for calculating the enhancement of
$J_{IMD}$ in an inhomogeneous 2D superconductor.
In Section \ref{sec:level3}, we give several analytical results for
the relevant enhancements and for $J_{IMD}$, 
based on the analogy to a nonlinear dielectric composite. Section
\ref{sec:level4} presents a concluding discussion.

\section{\label{sec:level2}Formalism}

\subsection{Intermodulation Coefficients from Ginzburg-Landau
Theory}

We begin by expressing $J_{IMD}$ in terms of
coefficients of the Ginzburg-Landau free energy density $F$.  
In the absence of a vector potential ${\bf A}$, $F$ takes the
standard form
\begin{equation}
F = a|\psi|^2 + \frac{b}{2}|\psi|^4 + C|{\bf \nabla}\psi|^2,
\label{eq:f}
\end{equation}
where $a$, $b$, and $C$ are appropriate constants, and $\psi$ is the
complex position-dependent Ginzburg-Landau wave function. 
For the present problem, we will eventually assume that all
three constants are functions of position. We also write $C =
\hbar^2/(2m^*)$, where $m^*$ is a quantity with dimensions of mass.

The local supercurrent density takes the usual quantum-mechanical
form
\begin{equation}
{\bf J} = \frac{e^*}{2m^*}\left[\psi^*(-i\hbar{\bf \nabla})\psi +
c.c.\right],
\end{equation}
where $e^*$ is the charge of a Cooper pair and `$c.c.$' denotes
a complex conjugate. We write $\psi = |\psi|\exp(i\phi)$ and
initially assume that $\phi$, but not $|\psi|$, is
position-dependent, so that
\begin{equation}
{\bf J} = \frac{e^*}{m^*}|\psi|^2(\hbar{\bf \nabla}\phi).
\label{eq:j}
\end{equation}
In the limit of very small current density, the wave function is
found by minimizing the free energy with respect to $|\psi|^2$,
which gives the standard expression $|\psi|^2 = -a/b$.  This
quantity is positive if $a < 0$.

If there is a finite phase gradient, $F$ takes the form
\begin{equation}
F = a|\psi|^2 + \frac{b}{2}|\psi|^4 +
\frac{\hbar^2}{2m^*}|\psi|^2|{\bf \nabla}\phi|^2.
\end{equation}
Minimizing with respect to the modulus of the wave function at fixed
${\bf \nabla}\phi$, we find that
\begin{equation}
|\psi|^2 = -\frac{\left[a + (\hbar^2/(2m^*))|{\bf
\nabla}\phi|^2\right]}{b}. 
\label{eq:psi1}
\end{equation}
The corresponding current density takes the form
\begin{equation}
{\bf J} = -\frac{e^*}{m^*}\left[\frac{a}{b}(\hbar{\bf \nabla}\phi) +
\frac{\hbar^2}{2m^* b}|{\bf \nabla}\phi|^2\hbar{\bf
\nabla}\phi\right]. \label{eq:j1}
\end{equation}

In the above discussion, we have assumed that $|\psi|$ is
position-independent, so that ${\bf \nabla}|\psi| = 0$.  Even if
${\bf \nabla}|\psi| \neq 0$, Eq.\ (\ref{eq:j}) for ${\bf J}$ remains
valid.  However, there is an extra term in the free energy density;
so Eq.\ (\ref{eq:psi1}), and hence Eq.\ (\ref{eq:j1}), do not hold
exactly. Nevertheless, we shall assume that the most important
effects of inhomogeneity can be included by writing
\begin{equation}
{\bf J} = K_1({\bf x}){\bf \nabla}\phi\left(1 - K_2({\bf x})|{\bf
\nabla}\phi|^2\right), 
\label{eq:j2}
\end{equation}
with appropriate coefficients $K_1({\bf x})$ and $K_2({\bf x})$.

We now show that Eq.\ (\ref{eq:j1}) (for a uniform superconductor)
contains the intermodulation phenomenon of interest.  First, we
rewrite Eq.\ (\ref{eq:j1}) for a uniform superconductor as
\begin{equation}
{\bf J} = K_1{\bf \nabla}\phi(1 - K_2|{\bf \nabla}\phi|^2),
\label{eq:j3}
\end{equation}
where $K_1$ and $K_2$ are related to the original $a$, $b$, and
$m^*$.   If ${\bf A} \neq 0$, in order for the
gradient to remain gauge-invariant, we must make the replacement $
-i\hbar{\bf \nabla} \rightarrow -i\hbar{\bf \nabla}-(e^*/c){\bf A}$,
or equivalently
\begin{equation}
{\bf \nabla}\phi \rightarrow {\bf \nabla}\phi - \frac{e^*}{\hbar
c}{\bf A}.
\end{equation}
Thus, if ${\bf A} \neq 0$ but the phase is uniform, we must rewrite
Eq.\ (\ref{eq:j3}) as
\begin{equation}
{\bf J} = -\frac{e^* K_1}{\hbar c}{\bf A}\left[1 -
\frac{e^{*2}K_2}{\hbar^2c^2}|{\bf A}|^2\right].
\end{equation}
To order $A^3$ (or $J^3$), we can replace $|{\bf A}|^2$ on the
right-hand side of this expression by $[\hbar c/(e^*K_1)]^2|{\bf
J}|^2$, which gives
\begin{equation}
{\bf J} = -\frac{e^* K_1}{\hbar c}{\bf A}\left[1 - \frac{
J^2}{J_{IMD}^2}\right], 
\label{eq:imod}
\end{equation}
where
\begin{equation}
J_{IMD}^2 = \frac{K_1^2}{K_2}. 
\label{eq:jimd}
\end{equation}

Finally, we show that Eq.\ (\ref{eq:imod}) implies a
current-dependent penetration depth.   To see this, we first take
the curl of Eq.\ (\ref{eq:imod}) in the low current limit to get
\begin{equation}
{\bf \nabla} \times {\bf J} = -\alpha^\prime {\bf B},
\end{equation}
where $\alpha^\prime = e^*K_1/(\hbar c)$.   When this equation is
combined with Ampere's law, ${\bf \nabla}\times{\bf B} = 4\pi {\bf
J}/c$, we get $\nabla^2{\bf J} - [1/\lambda^2(T,0)]{\bf J} = 0$,
where
\begin{equation}
\frac{1}{\lambda^2(T, 0)} = \frac{4\pi e^*K_1}{\hbar c^2},
\end{equation}
and $\lambda(T, 0)$ is the zero-current penetration depth at
temperature $T$. Thus finally
\begin{equation}
{\bf J} = -\frac{c}{4\pi\lambda^2(T, 0)}{\bf A}\left[1 -
\frac{J^2}{J_{IMD}^2}\right]. 
\label{eq:ja}
\end{equation}
Eq.\ (\ref{eq:ja}) has the form ${\bf J} = -\mu(T,J){\bf A}$, where
$\mu(T,J) = c/[4\pi\lambda^2(T,J)]$, $\lambda(T,J)$ being the
temperature- and current-dependent penetration depth. Thus, Eq.\
(\ref{eq:ja}) is equivalent to
\begin{equation}
\frac{1}{\lambda^2(T,J)} = \frac{1}{\lambda^2(T,0)}\left[1 -
\frac{J^2}{J_{IMD}^2}\right].
\end{equation}
To order $J^2$, this result is equivalent to Eq.\ (7) of Ref.\
\cite{dahm1} \cite{note}. 

\subsection{Estimate of Ginzburg-Landau Parameters for Cuprate
Superconductors}

Within the Ginzburg-Landau formalism, the penetration depth is
related to the order parameter $|\psi|$ by
\begin{equation}
\frac{1}{\lambda(T,0)} = \left[\frac{4\pi
e^{*2}|\psi|^2}{m^*c^2}\right]^{1/2},
\end{equation}
and $|\psi|^2 = -a/b$, provided
that $a < 0$.  Taking $a = -\alpha(1 - T/T_{c0})$, where $\alpha
> 0$ and 
$T_{c0}$ is the mean-field transition temperature, we obtain
\begin{equation}
|\psi|^2 = \alpha\left(1 - T/T_{c0}\right)/b 
\label{eq:psi2}
\end{equation}
and
\begin{equation}
\frac{1}{\lambda^2(T,0)} = \frac{4\pi e^{*2}\alpha(1-T/T_{c0})}
{m^*c^2b}.
\label{eq:lambda2}
\end{equation}

In Eq.\ (\ref{eq:lambda2}),
$\psi$ which has dimensions of a wave function. Lacking
an accepted microscopic theory for the high-T$_c$
cuprates, we may estimate $\alpha$ and $b$ using BCS theory, as
discussed, for example, by de Gennes \cite{gennes}. The result is
\begin{equation}
\alpha = \frac{\hbar^2}{2m^*\xi_0^2}  
\label{eq:alpha}
\end{equation}
and
\begin{equation}
b = \frac{\hbar^4}{4m^{*2}\xi_0^4}\frac{0.1066}{N(0)(k_BT_{c0})^2}.
\label{eq:b}
\end{equation}
Here, $N(0)$ is the single-particle density of states at the Fermi
energy (measured in states per unit energy per unit volume), and
$\xi_0$ is a temperature-independent coherence length.  The 
penetration depth is then found to be determined by the equation
\begin{equation}
\frac{1}{\lambda^2(T,0)} = \frac{32\pi e^2}{\hbar^2
c^2}N(0)\xi_0^2\Delta^2(T), 
\label{eq:lambdelta}
\end{equation}
where $\Delta(T)$ is the equilibrium value of the energy gap, given
by the equation
\begin{equation}
\Delta^2(T) = 9.38 k_B^2T_{c0}(T_{c0} - T).
\end{equation}

Now according to experiments \cite{davis1,davis2}, in the small
gap regions, BSCCO-2212 has a sizable superfluid density, whereas in
the large-gap regions, the superfluid density is small or zero.  If
we interpret $1/\lambda^2(T,0)$ as proportional to the local
superfluid density, then this experimental result implies that
$1/\lambda^2(T,0)$ should vary {\em inversely} with $\Delta(T)$. In
order for this to be consistent with Eq.\ (\ref{eq:lambdelta}), the
quantity $N(0)\xi_0^2\Delta^2(T)$ should vary inversely with
$\Delta(T)$.  We therefore assume that $1/\lambda^2(T,0) \propto
\Delta^{-x}(T)$, where $x >0$ is some number which characterizes
BSCCO-2212.  While this is a highly oversimplified model, it does
suggest how $J_{IMD}$ is influenced by the inhomogeneity
of the high-T$_c$ layer.

\section{\label{sec:level3}Model Calculations}

\subsection{Analogy to a Composite Dielectric Medium with a Cubic
Nonlinearity}

We now apply the above results to calculate $J_{IMD}$ 
for several models of
inhomogeneous superconducting layers.  In all cases, we attempt to
choose the layer properties to resemble
those reported in experiments on BSCCO-2212.
Our goal is to solve Eq.\ (\ref{eq:j2}) for ${\bf J}({\bf x})$ and
$\phi({\bf x})$ for some prescribed inhomogeneous superconductor. We
assume that $K_{1}({\bf x})$ and $K_{2}({\bf x})$ are specified, but
random.

The present problem is {\it formally} equivalent
to a randomly inhomogeneous dielectric with a cubic
nonlinearity \cite{zhang,stroud1}.  In the latter problem, the electric 
field ${\bf E}$ and electric displacement ${\bf D}$ are related by 
\begin{eqnarray}
{\bf D}({\bf x}) & = & \epsilon({\bf x}){\bf E}({\bf x}) +
\chi({\bf x})|{\bf E}({\bf x})|^2{\bf E}({\bf x}), \nonumber \\
{\bf \nabla} \times {\bf E} & = & 0, \nonumber \\
{\bf \nabla}\cdot{\bf D} & = & 0. 
\label{eq:dielec}
\end{eqnarray}
For the intermodulation problem, the analogous equations are Eq.\ 
(\ref{eq:j2}), supplemented by the steady-state charge conservation 
condition ${\bf \nabla} \cdot {\bf J} = 0$.
Thus, $-{\bf \nabla}\phi$ plays the role of ${\bf E}$
in the intermodulation problem, and the phase $\phi$ plays the
role of the scalar potential. The quantities $K_1({\bf x})$ and
$K_1({\bf x})K_2({\bf x})$ are analogous to the linear dielectric
function $\epsilon({\bf x})$ and cubic nonlinear susceptibility
$\chi({\bf x})$.  The quantity $-{\bf \nabla}\phi$ is, of course,
curl-free like ${\bf E}$ in an electrostatic problem. Thus,
we are again connecting a divergence-free field to a curl-free
field. To treat the intermodulation problem, therefore, we can use
all the formal results previously derived for an inhomogeneous
nonlinear dielectric, which we now briefly review.

For a material described by Eqs.\ (\ref{eq:dielec}), two useful
coefficients are the {\it effective linear dielectric function}
$\epsilon_e$ and the {\it effective cubic nonlinear susceptibility}
$\chi_e$.  These quantities are defined in terms of
the space-averaged electric field $\langle {\bf E}\rangle$ and
displacement $\langle {\bf D}\rangle$ by
\begin{equation}
\langle {\bf D}\rangle = \epsilon_e \langle {\bf E}\rangle + \chi_e
|\langle {\bf E}\rangle|^2\langle{\bf E}\rangle.
\end{equation}
As shown in Ref.\ \cite{stroud1}, $\chi_e$ can be expressed as an
average over the fourth power of the electric field in the
associated {\it linear} composite.  That is, if ${\bf E}_{\mathrm{lin}}
({\bf x})$ is the electric field in a composite described by the linear
relation ${\bf D}({\bf x}) = \epsilon({\bf x}){\bf E}({\bf x})$,
then $\chi_e$ is given by
\begin{equation}
\chi_e E_0^4 = \langle \chi({\bf x})|{\bf E}_{\mathrm{lin}}({\bf
x})|^4\rangle, 
\label{eq:moment}
\end{equation}
where $E_0$ is the applied electric field.   If the composite medium
has $n$ components, the $i$th of which has nonlinear susceptibility
$\chi_i$, then Eq.\ (\ref{eq:moment}) can be rewritten as
\begin{equation}
\chi_e = \sum_i p_i \chi_i e_i, 
\label{eq:moment1}
\end{equation}
where $p_i$ is the volume fraction of the $i$th component, $e_i$ is
an enhancement factor given by
\begin{equation}
e_i = \frac{\langle |{\bf E}({\bf x})|^4\rangle_{i,\mathrm{lin}}}{E_0^4},
\label{eq:ei}
\end{equation}
and $\langle \cdots \rangle_{i,\mathrm{lin}}$ means a space-average
within the $i$th component in the related linear medium.  Thus, $e_i$
describes how much the fourth power of the electric field is
enhanced in the $i$th component in the linear limit.

The moments $e_i$ are difficult to compute exactly, except in a few
very simple geometries.  We have therefore chosen to make a
decoupling approximation \cite{stroud2,zeng}, specified by
\begin{equation}
\langle |{\bf E}({\bf x})|^4\rangle_{i,\mathrm{lin}} \approx \langle
|{\bf E}({\bf x})|^2\rangle_{i,\mathrm{lin}}^2. \label{eq:decouple}
\end{equation}
Clearly, the decoupling approximation (\ref{eq:decouple}) will be most accurate 
if the fluctuations $\langle |{\bf E}({\bf x})|^4\rangle_{i,\mathrm{lin}} -
\langle |{\bf E}({\bf x})|^2\rangle_{i,\mathrm{lin}}^2$ within the $i$th component
are small compared to $\langle|{\bf E}({\bf x})|^4\rangle_{i,\mathrm{lin}}$ itself 
\cite{zeng}. This assumption is most likely to be accurate in geometries such that 
$|{\bf E}({\bf x})|$ is uniform in each of the nonlinear components, but will be 
less accurate when the fluctuations are large.  For example, in the so-called 
Hashin-Shtrikman geometry, in which one of the two components is embedded in 
the other, these fluctuations are small in the embedded component, and hence this 
approximation will be excellent if only the embedded component is nonlinear. 
However, in a composite near a percolation threshold, the fluctuations will be 
large and this approximation will be less accurate.  

If we make the approximation (\ref{eq:decouple}), we can express 
$\langle |{\bf E}({\bf x})|^2\rangle_{i,\mathrm{lin}}$ exactly in terms of the 
effective linear dielectric function $\epsilon_e$ through the relation
\begin{equation}
\frac{p_i\langle |{\bf E}({\bf x})|^2\rangle_{i,\mathrm{lin}}}{E_0^2} =
\frac{\partial\epsilon_e}{\partial\epsilon_i} \equiv F_i.
\label{eq:fi}
\end{equation}
Here, $\partial\epsilon_e/\partial\epsilon_i$ is the partial
derivative of $\epsilon_e$ with respect to $\epsilon_i$, at constant
$\epsilon_j$ ($j\neq i$) and constant volume fractions $p_j$. Given
a simple analytical approximation for $\epsilon_e$, these
derivatives can be easily computed in closed form, thus yielding a
simple analytical approximation for $\chi_e$ as
\begin{equation}
\chi_{e}=\sum_{i}\chi_{i}F_{i}^{2}/p_{i}
\label{eq:chi}
\end{equation}
with $e_{i}=F_{i}^{2}/p_{i}^{2}$.
We will use this approach, combined with the analogy described above,
to obtain approximations for the intermodulation coefficient in an
inhomogeneous superconducting layer.

In the present work, we use two different approximation methods to
calculate $\epsilon_e$: the effective-medium
approximation (EMA) \cite{bruggeman,landauer}, and the Maxwell-Garnett
approximation (MGA) \cite{landauer}. The EMA is
suitable for a binary composite with symmetrically distributed 
components, so that neither can be viewed as the inclusion or the 
host \cite{stroud2}.   In this
case, if the components are isotropic, $\epsilon_e$ satisfies the
quadratic equation
\begin{equation}
p_{A}\,\frac{\epsilon_{A} -
\epsilon_{e}}{\epsilon_{e}+g(\epsilon_{A} -\epsilon_{e})} +
(1-p_{A})\,\frac{\epsilon_{B}-\epsilon_{e}}
{\epsilon_{e}+g(\epsilon_{B} - \epsilon_{e})} = 0. 
\label{eq:ema}
\end{equation}
Here $p_{A}$ is the volume fraction of the component $A$,
$\epsilon_{A}$ and $\epsilon_{B}$ are the dielectric functions of
the components $A$ and $B$, respectively, and $g$ is a
`depolarization factor': $g=1/2$ in two dimensions (2D) and
$g=1/3$ in 3D. The physically meaningful solution of Eq.\
(\ref{eq:ema}) is the root which varies continuously with $p_A$, 
approaches the correct limits at $p_A = 0$ and $1$, and has
a nonnegative imaginary part when $\epsilon_A$ and $\epsilon_B$ are 
complex \cite{stroud2}.

The MGA is more suitable to a binary
composite where one component can be regarded as a host in which the
other is embedded \cite{zeng}. When the host material is
isotropic and linear, the effective dielectric function of the
composite takes the form
\begin{equation}
\epsilon_{e} = \epsilon_{h}\left[1 + \frac{p_{i}\,(\epsilon_{i}
- \epsilon_{h})}{(1-p_{i})[\epsilon_{h}(1-g) + \epsilon_{i}\,g]
+ p_{i}\,\epsilon_{h}}\right],
\label{eq:mga}
\end{equation}
where $p_{i}$ is the volume fraction of the inclusion, $\epsilon_{i}$
and $\epsilon_{h}$ are the dielectric functions of the inclusion and
the host, respectively, and $g$ is again the depolarization factor.

\subsection{Application to an Inhomogeneous Superconductor}

We can readily use the above analogy to compute the effective
nonlinear coefficients for an inhomogeneous superconducting layer.
We consider a superconducting layer comprised of two ``components,''
$A$ and $B$, with areal fractions $p_A$ and $p_B = 1 - p_A$, which
have two different energy gaps.  The two components are both
intrinsic to the given sample, in the sense that they are not caused
by the introduction of non-superconducting impurities.  A
realistic sample of BSCCO-2212 probably has a continuous
distribution of gaps, but we make this simplification for
computational convenience.  

The effective cubic nonlinear coefficient $(K_1K_2)_e$ takes the
form
\begin{equation}
(K_{1}K_{2})_{e}  =  (K_{1}K_{2})_{A}p_{A}e_{A}
+ (K_{1}K_{2})_{B}p_{B}e_{B}.
\label{eq:k1k2}
\end{equation}
$J_{IMD}$ in Eq.\
(\ref{eq:jimd}) thus becomes
\begin{equation}
J_{IMD} = \frac{K_{1e}}{K_{2e}^{1/2}} =
\frac{(K_{1}K_{2})_{e}}{K_{2e}^{3/2}}. 
\label{eq:jimd2}
\end{equation}

To apply the present formalism, we need to find suitable $K_{1}$ and $K_{2}$
values. From Eqs.\ (\ref{eq:j1}) and (\ref{eq:j3}), $K_{1} =
-(a/b)(\hbar e^*/m^*)$ and $K_{1}K_{2} = \hbar^3 e^*/(2m^{*2} b)$,
giving $K_{2} = -\hbar^{2}/(2m^{*}a)$. Using $a = \alpha(t-1)$, 
where $t = T/T_{c0}$, and taking $\alpha$ and $b$ from Eqs.\ 
(\ref{eq:lambda2}) and (\ref{eq:alpha}), we find
\begin{equation}
K_{1} = -\frac{a}{b}\frac{\hbar e^*}{m^*} = \frac{\hbar c^2}
{4\pi e^* \lambda^2(t,0)} 
\label{eq:klambda}
\end{equation}
and
\begin{equation}
K_{2} = \frac{\hbar^2}{2m^* |a|} = \xi^{2}(t).
\end{equation}

In typical cuprate superconductors, $\xi(t=0) \sim 15 \:\mbox{\AA}$
and $\lambda(t=0,J=0) \sim 1500 \:\mbox{\AA}$; so
\begin{equation}
K_{2}=(15 \:\mbox{\AA})^{2}.
\end{equation}
To estimate the values of $K_{1}$, we first assume that
\begin{equation}
\frac{1}{\lambda^2(T,0)} \propto [\Delta(T)]^{-x}. 
\label{eq:prop}
\end{equation}
This assumption embodies the experimental observation that the
superfluid density is relatively large in regions where the gap is
relatively small.  Thus, it is simply an effort to include relevant
experimental features in the model, without attempting to explain
them.  In the model calculations below, we consider two different values
of $x$, in order to see how this value affects the calculated
$J_{IMD}$.

Eqs.\ (\ref{eq:klambda}) and (\ref{eq:prop}) can be combined with 
experiment to get a rough estimate of $K_1$.
According to Ref.\ \cite{davis1}, $\Delta(T)$ ranges from 
$25\:\mathrm{meV}$ to $65\:\mathrm{meV}$ in 
optimally doped BSCCO-2212 at low $T$.  
We assume that the mean value, $45\:\mathrm{meV}$, corresponds to 
$\lambda(0,0) = 1500\:\mbox{\AA}$.
This fixes the proportionality constant in Eq.\
(\ref{eq:prop}).  Using this proportionality constant and Eq.\
(\ref{eq:klambda}), we get
\begin{equation}
K_1 = 3.49\times 10^{11}\left(\frac{\Delta(0)}{45\:\mathrm{meV}}
\right)^{-x}\frac{\mathrm{esu}}{\mathrm{cm}\cdot\mathrm{s}}. 
\label{eq:prop1}
\end{equation}

We first assume that $x = 1$.  Eq.\
(\ref{eq:prop1}) implies that 
$\Delta(0) = 25\:\mathrm{meV}$ and $65\:\mathrm{meV}$ correspond 
respectively to $K_{1A} = 6.28 \times
10^{11}\:\mathrm{esu}/(\mathrm{cm}\cdot\mathrm{s})$ and 
$K_{1B} = 2.42 \times 10^{11}\:
\mathrm{esu}/(\mathrm{cm}\cdot\mathrm{s})$.  For $K_{2A}$ and $K_{2B}$, we
have no definite information from experiment.  We therefore assume simply that
$K_{2A} = K_{2B}$.

\begin{figure}[h]
\begin{center}
\includegraphics[width=\textwidth]{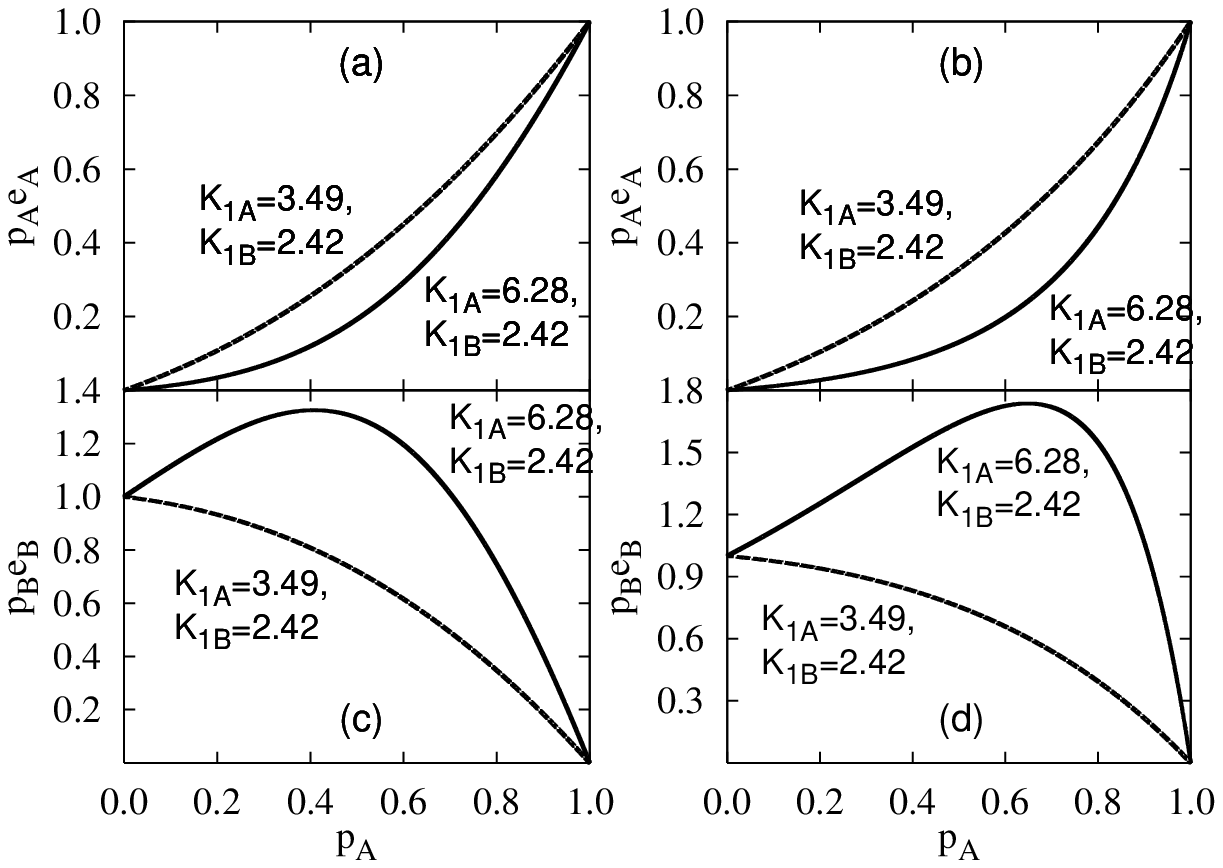}\\%
\end{center}
\caption{\label{fig:enhance1}Enhancement factors $p_Ae_A$ 
and $p_Be_B$ for a 2D inhomogeneous superconductor consisting 
of a binary composite with two different energy gaps, which have the 
ratio $K_{1A}/K_{1B}$. The effective-medium approximation
(EMA) is used in (a) and (c), while the Maxwell-Garnett
approximation (MGA) is used in (b) and (d), assuming that
$A$ is surrounded by $B$.  We use two different sets of $K_1$'s:
$K_{1A}=6.28\times 10^{11}\:\mathrm{esu}/(\mathrm{cm}
\cdot\mathrm{s})$, $K_{1B}=2.42\times 10^{11}\:\mathrm{esu}
/(\mathrm{cm}\cdot\mathrm{s})\,$ and $K_{1A}=3.49\times
10^{11}\:\mathrm{esu}/(\mathrm{cm}\cdot \mathrm{s})$,
$K_{1B}=2.42\times 10^{11}\:\mathrm{esu}/(\mathrm{cm}
\cdot\mathrm{s})$ as indicated in the legend. In this and
the following plots, the unit of $K_{1A}$ and $K_{1B}$ is
$10^{11}\,\mathrm{esu}/(\mathrm{cm}\cdot\mathrm{s})$.}
\end{figure}

\begin{figure}[h]
\begin{center}
\includegraphics[width=\textwidth]{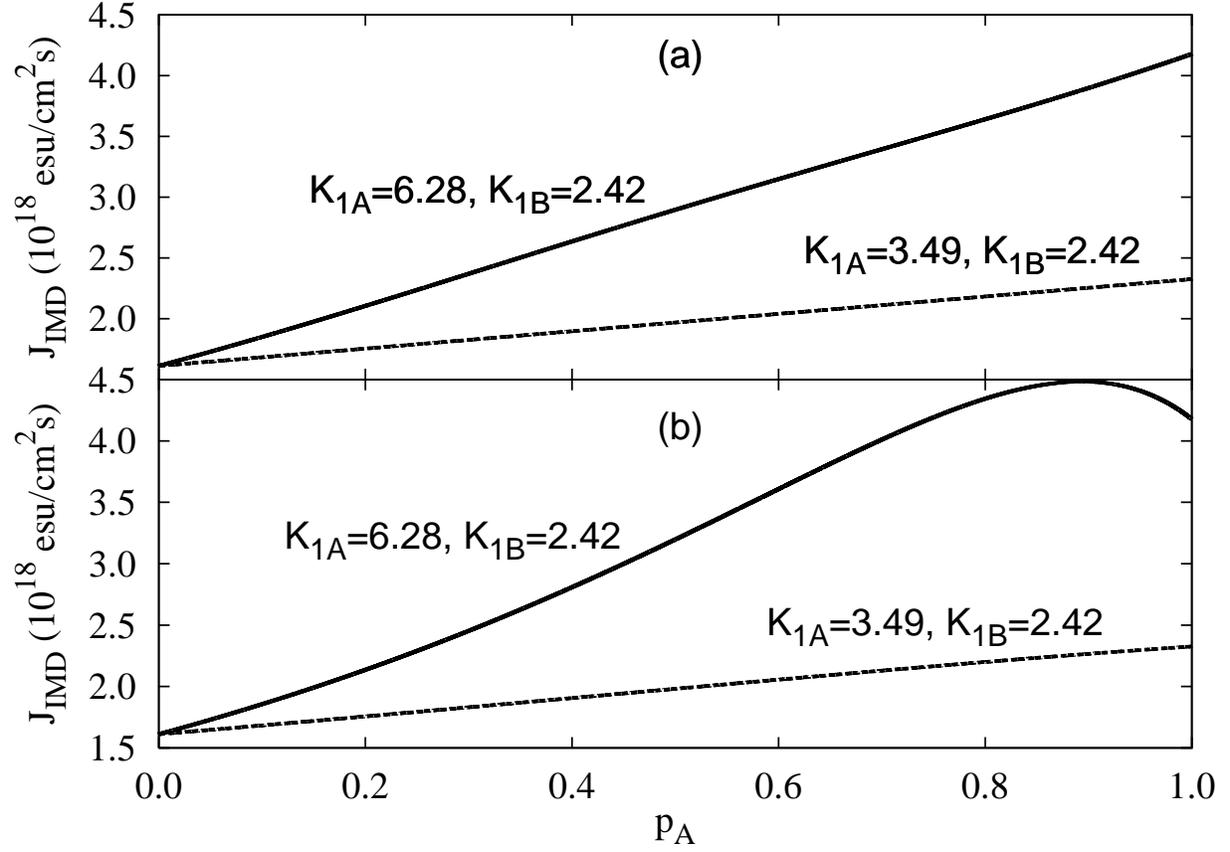}\\%
\end{center}
\caption{\label{fig:current1}Intermodulation critical supercurrent 
density $J_{IMD}$ for the 2D inhomogeneous superconductor shown in 
Fig.\ \ref{fig:enhance1}. The EMA method is used in (a), while the 
MGA method is used in (b), assuming that $A$ is surrounded by $B$. 
The two sets of $K_{1}$'s are the same as in Fig.\ \ref{fig:enhance1}. 
Note $10^{18}$ power scale on the $y$-axis.}
\end{figure}

In Figs.\ \ref{fig:enhance1} and \ref{fig:current1}, we show the
calculated enhancement factors $e_A$ and $e_B$ and the corresponding
intermodulation critical current density $J_{IMD}$ for these models, 
as functions of $p_A$.  In Figs.\ \ref{fig:enhance1}(a),
\ref{fig:enhance1}(c), and \ref{fig:current1}(a) we use the
EMA [Eq. (\ref{eq:ema})], while in Figs.\
\ref{fig:enhance1}(b), \ref{fig:enhance1}(d), and
\ref{fig:current1}(b) we use the MGA [Eq.\ (\ref{eq:mga})], with $B$ 
considered as the host. In both cases, we combine these approximations 
with the decoupling approximation [Eqs.\ (\ref{eq:decouple})--(\ref{eq:chi})], 
to obtain $J_{IMD}$.

Fig.\ \ref{fig:current1}(a) shows that $J_{IMD}$ increases linearly
with $p_A$ in the EMA.  As in Eqs.\ (\ref{eq:k1k2}) and
(\ref{eq:jimd2}), $J_{IMD}$ has contributions from the
nonlinearity of both components. While the enhancement factor $e_A$
is never larger than unity, $e_B$ can exceed
unity, depending on the value of $p_{A}$.  Thus, the nonlinearity of
$B$ has a larger influence on $J_{IMD}$ than that of $A$.  As a
result, $J_{IMD}$ behaves similarly to $e_B$, having a larger
enhancement for the larger $K_{1A}/K_{1B}$.
The MGA results differ little from the EMA results except for a
broad peak around $p_{A}=0.9$ for the larger ratio of
$K_{1A}/K_{1B}$. This peak results from the shift to higher values
of $p_A$ of both $p_Be_B$ and $p_Ae_A$, seen in the MGA results of
Fig.\ \ref{fig:enhance1}.

\begin{figure}[h]
\begin{center}
\includegraphics[width=\textwidth]{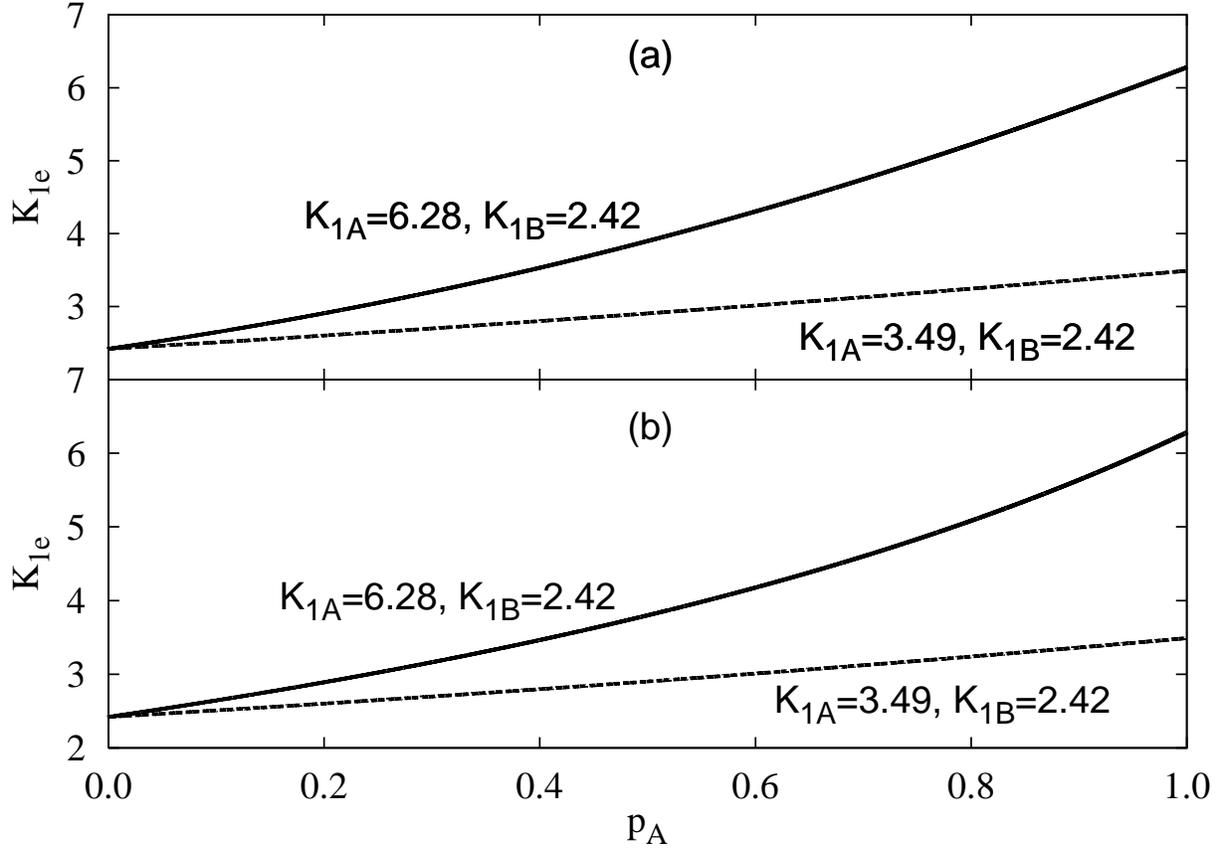}\\%
\end{center}
\caption{\label{fig:super1}Effective superfluid density 
$K_{1e}$ for a 2D composite having the same values of
$K_{1A}$ and $K_{1B}$ used in the calculations of Figs.\ 
\ref{fig:enhance1} and \ref{fig:current1}. (a) EMA method;  
(b) MGA method, taking B (the component with the smaller 
superfluid density) as the host.}
\end{figure}

We can also calculate the effective {\em linear} coefficients
$K_{1e}$ for these two models.  In the 2D EMA, $K_{1e}$ satisfies
\begin{equation}
p_A\frac{K_{1A}-K_{1e}}{K_{1A}+K_{1e}}+(1-p_A)\frac{K_{1B}-K_{1e}}
{K_{1B}+K_{1e}} = 0, 
\label{eq:emak1}
\end{equation}
while in the 2D MGA with $B$ considered as the host, we get
\begin{equation}
K_{1e} = K_{1B}\left[1 + \frac{2 p_A(K_{1A}-K_{1B})}{(1-p_A)
(K_{1A}+K_{1B})+2 p_A K_{1B}}\right].
\label{eq:mgak1} 
\end{equation} 
The $K_{1e}$'s are proportional to the effective superfluid 
densities (or the effective inverse-square penetration 
depths) of these 2D materials in the linear limit of very small applied 
currents. The values calculated from the EMA and MGA are shown in Figs.\ 
\ref{fig:super1}(a) and \ref{fig:super1}(b).  Both increase monotonically 
with increasing areal fraction of the small-gap component $A$. The MGA 
results differ very little from the EMA results. 

\begin{figure}[h]
\begin{center}
\includegraphics[width=\textwidth]{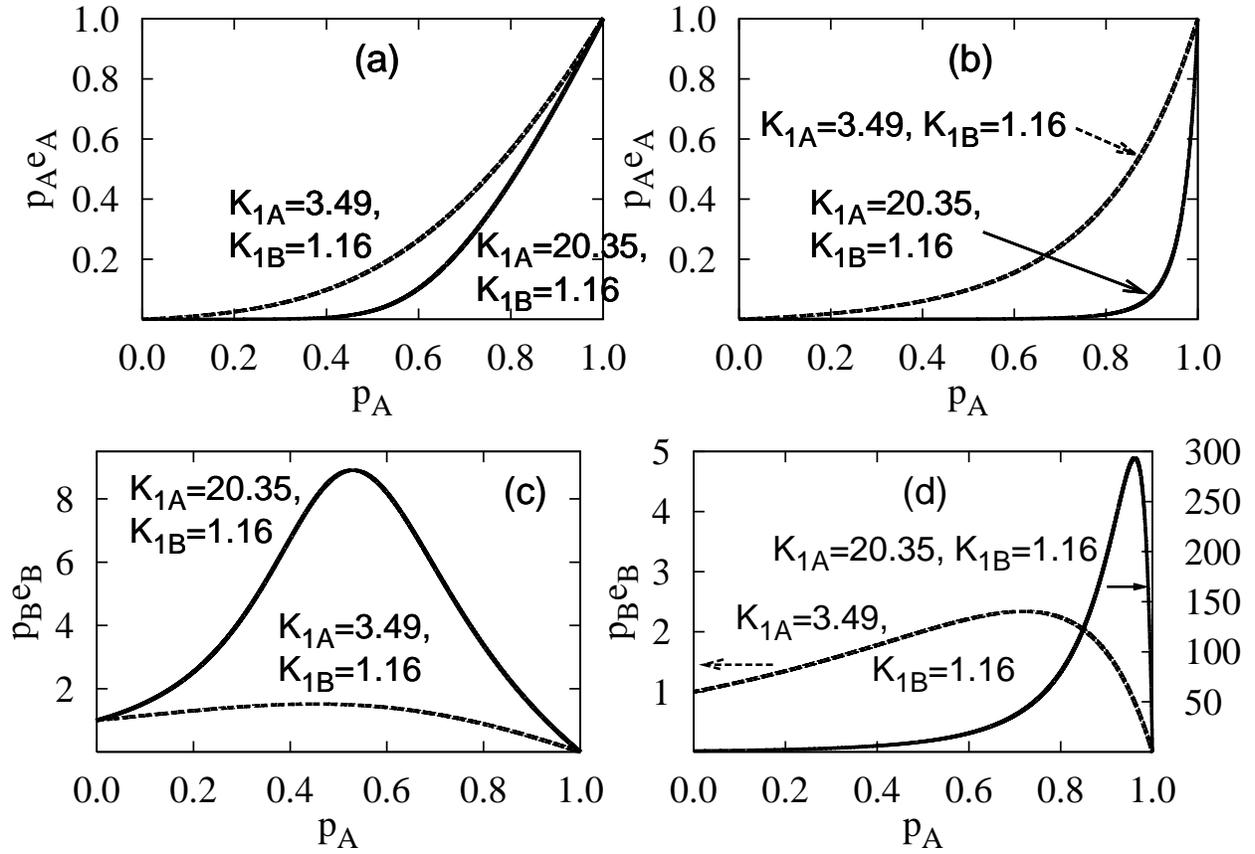}\\%
\end{center}
\caption{\label{fig:enhance2}Same as Fig.\ \ref{fig:enhance1}, 
except that $x=3$.  The two sets of $K_{1}$'s are 
$K_{1A}=20.35\times 10^{11}\:\mathrm{esu}/(\mathrm{cm}\cdot
\mathrm{s})$, $K_{1B}=1.16\times 10^{11}\:\mathrm{esu}
/(\mathrm{cm}\cdot\mathrm{s})\,$ and $K_{1A}=3.49\times 
10^{11}\:\mathrm{esu}/(\mathrm{cm}\cdot\mathrm{s})$, 
$K_{1B}=1.16\times 10^{11}\:\mathrm{esu}/(\mathrm{cm}
\cdot\mathrm{s})$. The full curve is scaled to the right 
axis while the dashed line to the left in (d).}
\end{figure}

\begin{figure}[h]
\begin{center}
\includegraphics[width=\textwidth]{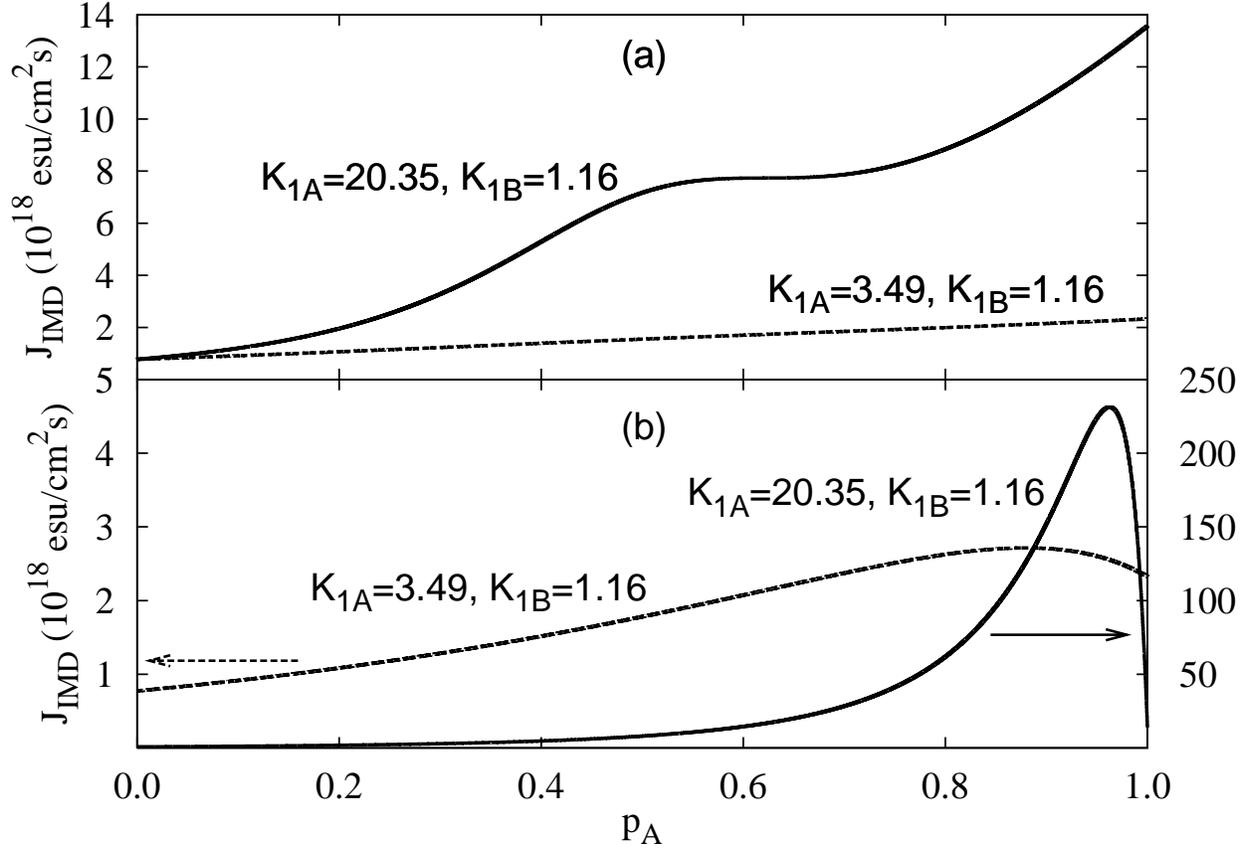}\\%
\end{center}
\caption{\label{fig:current2}Same as Fig.\ \ref{fig:current1}, 
except that $x=3$; so the two sets of $K_{1}$'s used are 
$K_{1A}=20.35\times 10^{11}\:\mathrm{esu}/(\mathrm{cm}
\cdot\mathrm{s})$, $K_{1B}=1.16\times 10^{11}\:\mathrm{esu}
/(\mathrm{cm}\cdot\mathrm{s})\,$ and $K_{1A}=3.49\times 
10^{11}\:\mathrm{esu}/(\mathrm{cm}\cdot\mathrm{s})$, 
$K_{1B}=1.16\times 10^{11}\:\mathrm{esu}/(\mathrm{cm}
\cdot\mathrm{s})$. The full curve is scaled to the right 
axis while the dashed line to the left in (b).}
\end{figure}

\begin{figure}[h]
\begin{center}
\includegraphics[width=\textwidth]{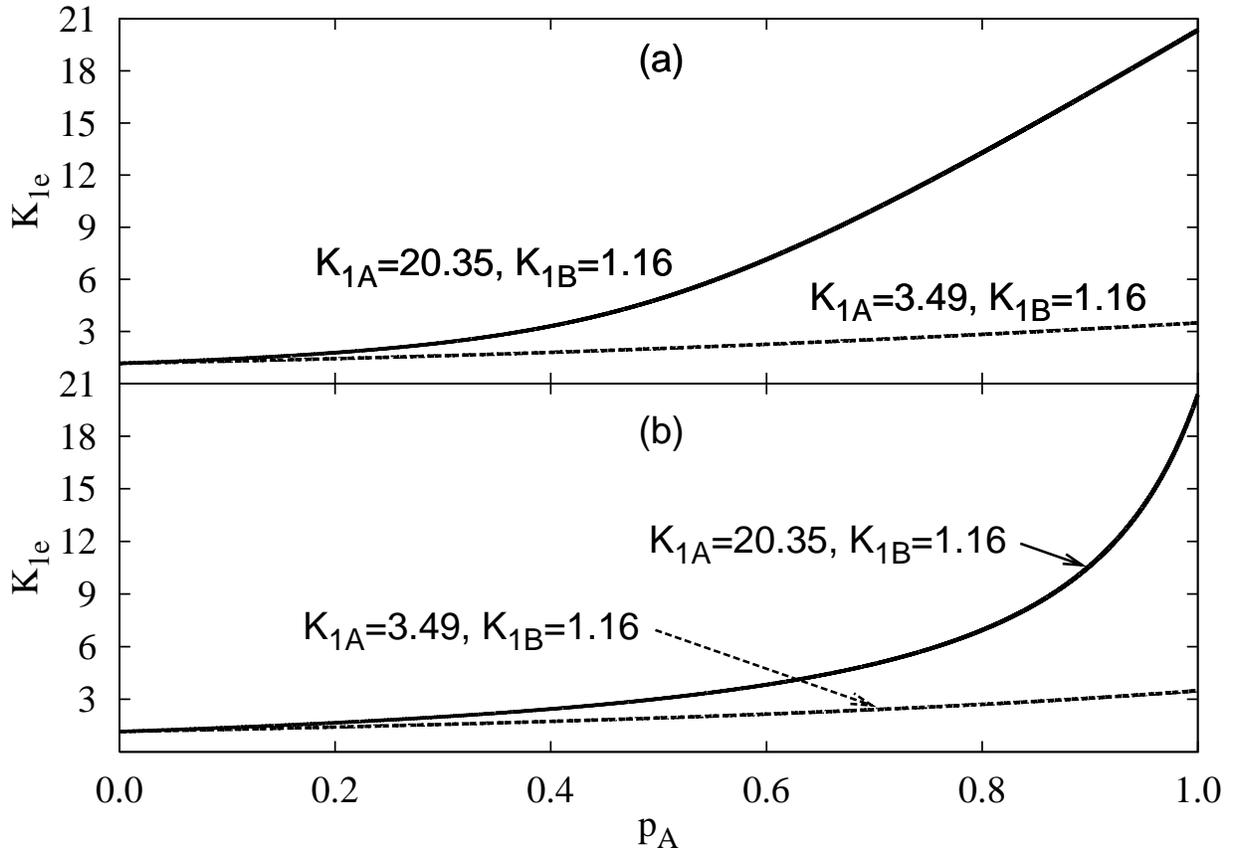}\\%
\end{center}
\caption{\label{fig:super2}Same as Fig.\ \ref{fig:super1} 
except that $x = 3$, corresponding to a much larger ratio 
of $K_{1A}/K_{1B}$.} 
\end{figure}

In Figs.\ \ref{fig:enhance2}, \ref{fig:current2}, and
\ref{fig:super2}, we show an analogous set of calculations, but with
$x = 3$.   We again assume a binary distribution of gaps, using the
same gaps as in Figs.\ \ref{fig:enhance1}--\ref{fig:super1}.  
Because of the larger $x$, the ratio $K_{1A}/K_{1B}$ is larger
than in Figs.\ \ref{fig:enhance1}--\ref{fig:super1}.  For $x = 3$, 
using the same proportionality
constant, we find that the gaps $\Delta(0) = 65\:\mathrm{meV}$ 
and $25\:\mathrm{meV}$ now correspond to
$K_{1} = 1.16\times 10^{11}\:\mathrm{esu}/(\mathrm{cm}\cdot\mathrm{s}) 
\equiv K_{1B}$, and $K_{1} =
20.35\times 10^{11}\:\mathrm{esu}/(\mathrm{cm}\cdot\mathrm{s}) \equiv K_{1A}$.

For $x=3$, for the larger ratio of $K_{1A}/K_{1B}$,
the enhancement factor $p_Be_B$ has clear peaks as a function of the 
areal fraction $p_A$. This behavior is shown in 
Figs.\ \ref{fig:enhance2}(c) and \ref{fig:enhance2}(d). The peak occurs 
at around the percolation threshold of $p_A = 0.5$ in the EMA results, 
but at around $p_{A}=0.95$ in the MGA results. In addition, $p_Be_B$ is 
dramatically larger ($\sim 100$) in the MGA than in the EMA, for the 
larger ratio of $K_{1A}/K_{1B}$. Note also 
that the EMA results are nearly symmetric about $p_A = 0.5$ while the 
MGA results are very asymmetric. There is also a large difference 
between results for the two gap ratios in the MGA results, but a 
smaller one in the EMA.  By contrast, $p_Ae_A$ is monotonic in either 
the EMA or the MGA. Since $J_{IMD}$ has two contributions, one from 
the enhancement of $A$ and the other from the enhancement of $B$, 
one expects that the behavior of $J_{IMD}$ in the MGA results 
will mirror the enhancement factor $p_Be_B$ because $p_Ae_A \ll p_Be_B$ 
for the larger ratio of $K_{1A}/K_{1B}$.

Fig.\ \ref{fig:current2} shows the behavior of $J_{IMD}$ for the
$K_{1}$'s shown in Fig.\ \ref{fig:enhance2}.  As expected, and as already
found for $x=1$, $J_{IMD}$ for $x=3$ generally follows the trend of $p_Be_B$.
In particular, because of the clear peak in $p_Be_B$, the EMA results show a weak
broad peak near the percolation threshold $p_c$ in $J_{IMD}$ for the
larger ratio of $K_{1A}/K_{1B}$.  The EMA and MGA results 
differ greatly for the larger ratio of 
$K_{1A}/K_{1B}$, not only in the shape of the curves but also in the 
magnitude of $J_{IMD}$.  In this case, the MGA results 
follow mostly the shape of the curve for $p_Be_B$. Although $J_{IMD}$ 
increases monotonically with $p_A$ in the EMA results, it drops sharply 
above $p_{A}=0.95$ for the larger ratio of $K_{1A}/K_{1B}$ in the 
MGA results. Overall, the $x=3$ case produces a much larger value 
of $J_{IMD}$ for the larger ratio of $K_{1A}/K_{1B}$. Thus, for a 
device requiring a large $J_{IMD}$, these results suggest that 
the best results would be obtained using an inhomogeneous 
superconductor with a large gap difference and a large $x$ in a 
Maxwell-Garnett geometry.

Fig.\ \ref{fig:super2} shows the effective superfluid densities $K_{1e}$
with $x=3$. For the smaller ratio of $K_{1A}/K_{1B}$, the EMA results
differ little from the MGA results as we found previously in Fig.\ \ref{fig:super1}. 
However, for the larger ratio of $K_{1A}/K_{1B}$, the two differ
significantly.

\begin{figure}[h]
\begin{center}
\includegraphics[width=\textwidth]{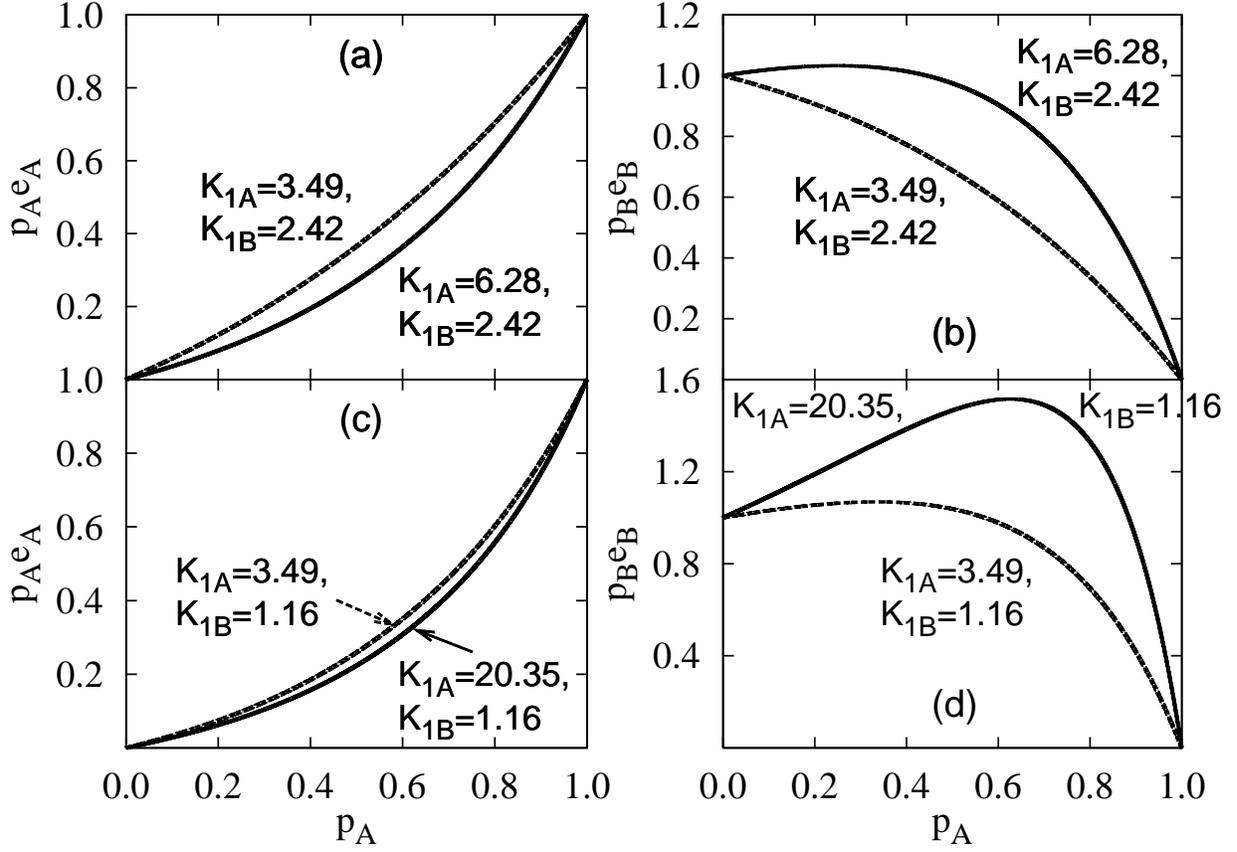}\\%
\end{center}
\caption{\label{fig:enhance3}Same as Fig.\ \ref{fig:enhance1}, 
except that the MGA method is used in all cases, assuming that 
$B$ is surrounded by $A$. The two sets of $K_{1}$'s are 
$K_{1A}=6.28\times 10^{11} \:\mathrm{esu}/(\mathrm{cm}\cdot
\mathrm{s})$, $K_{1B}=2.42\times 10^{11}\:\mathrm{esu}
/(\mathrm{cm}\cdot\mathrm{s})\,$ and $K_{1A}=3.49\times 
10^{11}\:\mathrm{esu}/(\mathrm{cm}\cdot\mathrm{s})$, 
$K_{1B}=2.42 \times 10^{11}\:\mathrm{esu}/(\mathrm{cm}\cdot
\mathrm{s})$ in (a) and (b), while they are $K_{1A}=20.35
\times 10^{11}\:\mathrm{esu}/(\mathrm{cm}\cdot\mathrm{s})$, 
$K_{1B}=1.16\times 10^{11}\:\mathrm{esu}/(\mathrm{cm}\cdot
\mathrm{s})\,$ and $K_{1A}=3.49\times 10^{11}\:\mathrm{esu}
/(\mathrm{cm}\cdot\mathrm{s})$, $K_{1B}=1.16\times 10^{11}\:
\mathrm{esu}/(\mathrm{cm}\cdot\mathrm{s})$ in (c) and (d).}
\end{figure}

\begin{figure}[h]
\begin{center}
\includegraphics[width=\textwidth]{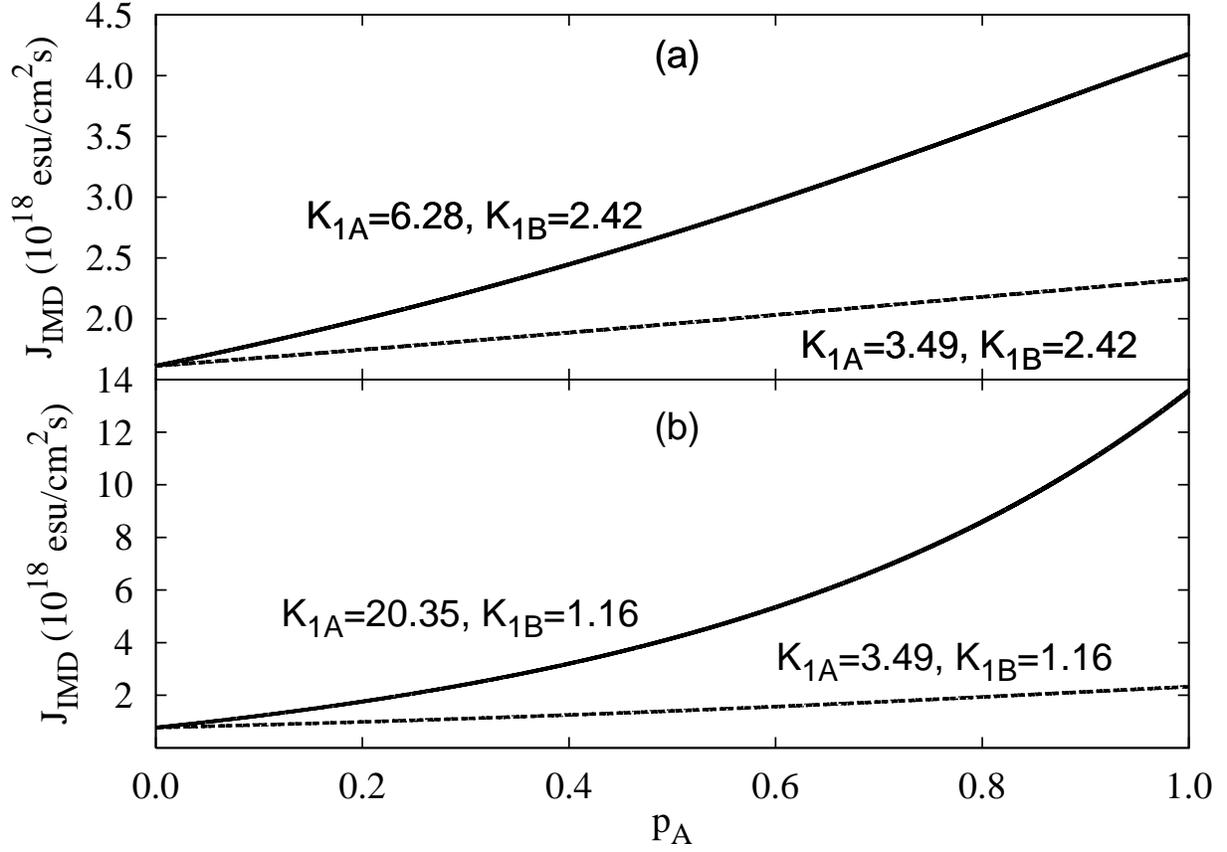}\\%
\end{center}
\caption{\label{fig:current3}Intermodulation critical 
supercurrent density $J_{IMD}$ for the 2D inhomogeneous
superconductor shown in Fig.\ \ref{fig:enhance3}. The MGA 
method is used in (a) and (b), assuming that $B$ is surrounded 
by $A$. The two sets of $K_{1}$'s used are the same as in
Fig.\ \ref{fig:enhance3}. Note $10^{18}$ power scale on the
$y$-axis.}
\end{figure}

\begin{figure}[h]
\begin{center}
\includegraphics[width=\textwidth]{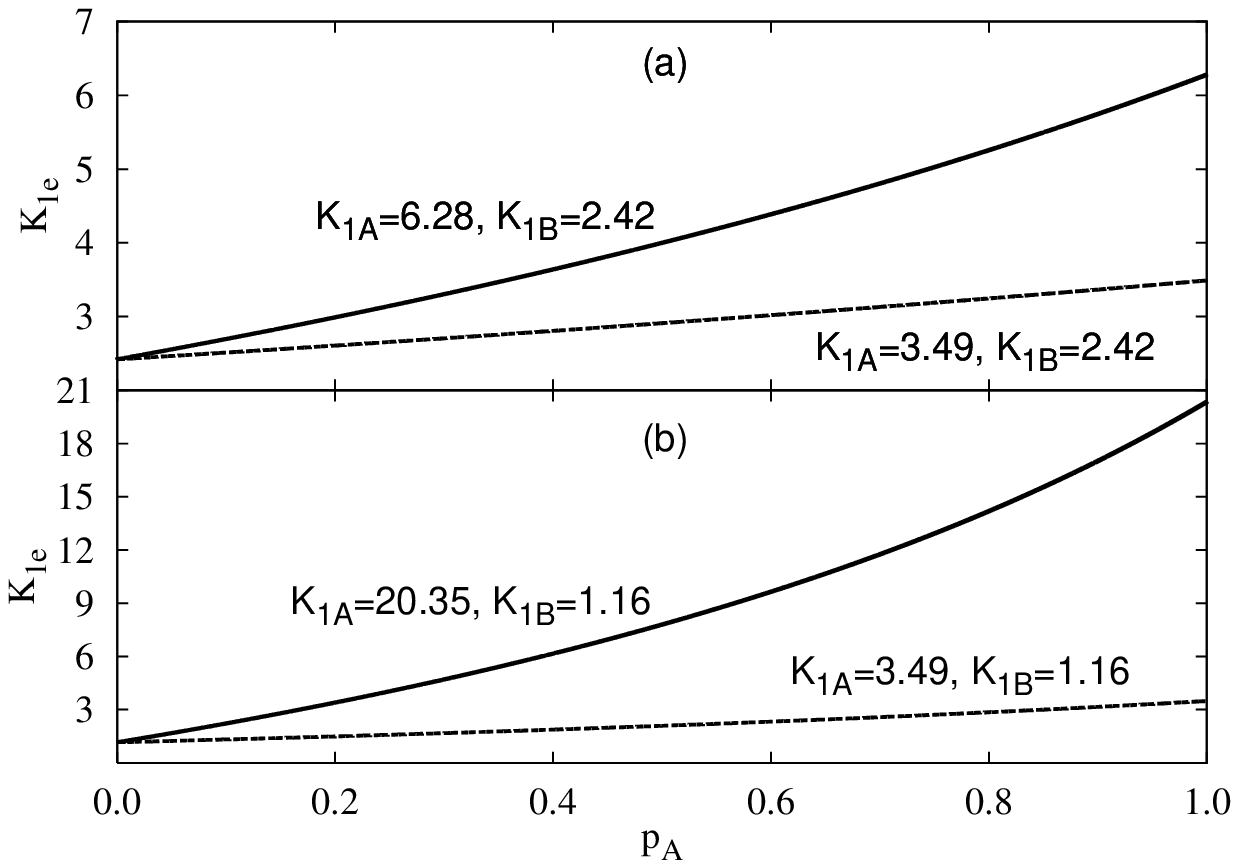}\\%
\end{center}
\caption{\label{fig:super3}Effective superfluid density 
$K_{1e}$ corresponding to Figs.\ \ref{fig:enhance3} and
\ref{fig:current3}.}
\end{figure}

There is no distinction between the inclusion and the host in the
EMA method, but there is in the MGA method.  In our MGA results thus
far, we have assumed that $A$ (the component with the smaller gap) is
the inclusion and that $B$ is the host.  We now consider the reverse
configuration, where $B$ is surrounded by $A$. Results for this
configuration are shown in Figs.\ \ref{fig:enhance3}, 
\ref{fig:current3}, and \ref{fig:super3}. The MGA results with $x=1$ 
shown in Figs.\ \ref{fig:enhance3}(a), \ref{fig:enhance3}(b), 
\ref{fig:current3}(a), and \ref{fig:super3}(a) are very similar to
the EMA results in Figs.\ \ref{fig:enhance1}(a), \ref{fig:enhance1}(c), 
\ref{fig:current1}(a), and \ref{fig:super1}(a), respectively; indeed, 
the results for $J_{IMD}$ and $K_{1e}$ are almost identical in the two
approximations.  For the larger value of $K_{1A}/K_{1B}$, the 
enhancement factor $p_Be_B$ is smaller in Fig.\ \ref{fig:enhance3}(b) 
than in Fig.\ \ref{fig:enhance1}(c).
In general, the results for this version of MGA do not show the
dramatically large increases in $J_{IMD}$ seen in Fig.\
\ref{fig:current2} for the larger ratio of $K_{1A}/K_{1B}$.
In Figs.\ \ref{fig:enhance3}(c), \ref{fig:enhance3}(d), 
\ref{fig:current3}(b), and \ref{fig:super3}(b) we show the same 
MGA calculations for $p_Ae_A$, $p_Be_B$, $J_{IMD}$, and $K_{1e}$ 
but with $x=3$. The behavior does not differ greatly from that 
seen in the cases with $x=1$, except for an increase in the 
magnitudes of $J_{IMD}$ and $K_{1e}$ for the larger ratio of 
$K_{1A}/K_{1B}$. By contrast, the MGA results for a $B$ host 
show some dramatic peaks for the larger ratio of $K_{1A}/K_{1B}$, 
as shown earlier.

\section{\label{sec:level4}Discussion}

In this paper, we have calculated the intermodulation critical
current $J_{IMD}$ in an inhomogeneous 2D superconductor
characterized by a binary distribution of energy gaps.  To carry out
this calculation, we used an analogy between the effective cubic
nonlinear response of an inhomogeneous superconductor and the
effective cubic nonlinear susceptibility of a composite dielectric
medium. Using this analogy, we can apply the formalism previously
developed to treat the nonlinear dielectric composite to the
inhomogeneous superconductor.  We found that the cubic nonlinear
response of the superconductor could be expressed in terms of the
cubic response of each ``component'' (i.e., energy gap), and two
enhancement factors, each describing the field and current
distribution in the related {\it linear} medium.

In order to simplify our calculations, we have assumed that the
superconducting layer has a binary distribution of energy gaps,
$\Delta_A$ and $\Delta_B$ (with $\Delta_B > \Delta_A$), and we have
considered three plausible topologies: ``effective-medium'' topology
($A$ and $B$ symmetrically distributed), and two ``Maxwell-Garnett''
topologies ($A$ embedded in $B$ and $B$ embedded in $A$).  We have
treated all three using a simple nonlinear decoupling approximation.

The results for $J_{IMD}$ are dramatically dependent on the assumed
topologies.  The EMA and the MGA with $B$ in $A$ give rather similar
results for ratios of $\Delta_B/\Delta_A$ close to unity, and only
modest enhancements of $J_{IMD}$ at any concentrations of $A$.
However, the MGA with $A$ (the component with the smaller gap) 
embedded in $B$ leads to huge
enhancements in $J_{IMD}$ compared to its value in either pure $A$
or pure $B$, provided that the two gap values are sufficiently 
different and that $x$ is large.

In view of these differences, it is of interest to compare our
results with the detailed measurements of Davis {\it et al.}
\cite{davis2}. These experiments do not provide results directly for
$J_{IMD}$. However, they do provide some hints about a
possible connection.  In particular, in experiments on as-grown
Ni-doped samples, Ni scattering resonances were
observed only in the regions where $\Delta < 50\:\mathrm{meV}$,
which were identified as superconducting regions, i.e.,
regions with superconducting phase coherence.  In our results, most
of the enhancement in $J_{IMD}$ comes from the enhancement factor
$e_B$, which corresponds to the component with a low energy gap. 
Therefore the regions of enhanced $J_{IMD}$ correspond to regions 
of small energy gap, and also regions of enhanced phase coherence 
according to the measurements of Ref.\ \cite{davis2}.

A striking feature of our results is the large difference between
the EMA results and the MGA results, especially for binary
composites with a large $x$ and very different energy gaps. Which 
of these approximations is the most correct?  In fact, there is 
not a single correct answer for all materials: the correct choice 
depends on the actual topology of the material of interest. In 
particular, we do not know, beforehand, whether the energy gaps in 
an inhomogeneous superconductor are distributed at random throughout 
the CuO$_{2}$ planes or whether regions with one magnitude of energy 
gap are preferentially surrounded by those of the other energy gap. 
This topology determines whether we should use the EMA or the MGA
approach.

In the experimental gap maps \cite{davis2}, the low-$\Delta$ regions
are surrounded by the high-$\Delta$ regions in the underdoped
BSCCO-2212 sample, but the high-$\Delta$ regions are surrounded by
the low-$\Delta$ regions in the slightly overdoped as-grown
BSCCO-2212 sample. Therefore, it appears that we can apply the MGA
method to both cases, but with the roles of inclusion and host
reversed in each case. But even this description of the
distribution may be a simplification of the true gap
distribution, which is probably continuous, not a discrete binary 
distribution. Ideally, we should consider such a continuous distribution 
of energy gaps.

The great sensitivity of $J_{IMD}$ to composite geometry, as found
in the present work, is not surprising, in view of earlier work on
transport in linear and nonlinear composite conductors and
dielectric media.  For example, it is well known that the critical
exponents describing transport, especially nonlinear transport, in
composite media are sensitive to features of the local geometry
\cite{rammal1,halperin,rammal2,tremblay}.  We speculate that,
depending on the precise nature of this geometry, $J_{IMD}$ either
diverges or goes to zero near a percolation threshold according to
an appropriate critical exponent.

In summary, we have presented a general formalism for calculating the 
intermodulation coefficient, and the corresponding intermodulation supercurrent 
density $J_{IMD}$, of an inhomogeneous superconductor. We have also given a 
simple way to calculate $J_{IMD}$ approximately in several geometries. Since 
such inhomogeneities are known to exist in many of the high-$T_c$ cuprate 
superconductors, this formalism is directly relevant for treating an important 
property of these materials. We find that the resulting $J_{IMD}$ is very 
sensitive to the exact spatial distribution of gaps within the inhomogeneous layer 
and thus may increase or decrease, depending on the topology. Our calculations 
show that one way to achieve a large $J_{IMD}$ is to have the component with 
the smaller gap and larger superfluid density embedded in the component with 
the opposite properties. This appears to be the topology seen in the underdoped
BSCCO-2212, which thus may be well suited for a material with a large $J_{IMD}$.

Finally, we comment briefly on possible device implications of the
present results.  A useful microstrip resonator would usually have a
minimum of intermodulation, or interference between different
frequencies \cite{remillard}.  To achieve this, one would probably
want a $J_{IMD}$ which is as large as possible, because this would
lead to $1/\lambda^2(T,J)$ which has the weakest dependence on
current. Surprisingly, we find here that $J_{IMD}$ can actually be 
{\it increased} in some inhomogeneous superconductors, provided that 
the topology is suitable. Thus the inhomogeneity which appears to be
unavoidable in the high-T$_c$ cuprates may even be an advantage in
constructing useful microwave devices.

\begin{acknowledgments}
This work was supported by NSF Grant DMR04-13395. We are grateful to
D. J. Bergman for helpful discussions.  Some of the calculations
were carried out on the P4 Cluster at the Ohio Supercomputer Center,
with the help of a grant of time.
\end{acknowledgments}

\newpage

\begin{center}
FIGURES
\end{center}

\vspace{0.1in}

\begin{thebibliography}{99}

\bibitem{bednorz} J. G. Bednorz and K. A. M\"{u}ller, Z. Physik B {\bf 64}, 189 (1986).

\bibitem{willemsen}  Balam A. Willemsen, T. Dahm, and D. J. Scalapino, Appl. Phys.
Lett. {\bf 71}, 3898 (1997).

\bibitem{dahm1} T. Dahm and D. J. Scalapino, Appl. Phys. Lett. {\bf 69}, 4248 (1996).

\bibitem{dahm2} T. Dahm and D. J. Scalapino, J. Appl. Phys. {\bf 81}, 2002 (1997).

\bibitem{yip} S. K. Yip and J. A. Sauls, Phys. Rev. Lett. {\bf 69}, 2264 (1992).

\bibitem{xu} D. Xu, S. K. Yip, and J. A. Sauls, Phys. Rev. B {\bf 51}, 16233 (1995).

\bibitem{stojkovic} Branko P. Stojkovi\'{c} and Oriol T. Valls, Phys. Rev. B {\bf 51},
6049 (1995).

\bibitem{mcdonald} J. McDonald, John R. Clem, and D. E. Oates, Phys. Rev. B {\bf 55},
11823 (1997).

\bibitem{davis1}  S. H. Pan, J. P. O'Neal, R. L. Badzey, C. Chamon, H. Ding,
J. R. Engelbrecht, Z. Wang, H. Eisaki, S. Uchida, A. K. Gupta, K.-W. Ng, E. W. Hudson,
K. M. Lang, and J. C. Davis, Nature {\bf 413}, 282 (2001).

\bibitem{davis2} K. M. Lang, V. Madhavan, J. E. Hoffman, E. W. Hudson,
H. Eisaki, S. Uchida, and J. C. Davis, Nature {\bf 415}, 412 (2002).

\bibitem{note}  We do not, in this paper, consider the dependence of
the coefficient $J_{IMD}$ on the angle $\theta$ between the current
and the $a$-axis of the CuO$_2$ plane, such as would be expected in
a superconductor having a $d_{x^2-y^2}$ order parameter.

\bibitem{gennes} P. G. de Gennes, \emph{Superconductivity of Metals and Alloys},
W. A. Benjamin, New York (1966), pp. 173--180; reprinted by Addison-Wesley (1989);
reprinted by Westview (1999).

\bibitem{zhang} X. Zhang and D. Stroud, Phys. Rev. B {\bf 49}, 944 (1994).

\bibitem{stroud1} D. Stroud and P. M. Hui, Phys. Rev. B {\bf 37}, 8719 (1988).

\bibitem{stroud2} D. Stroud and Van E. Wood, J. Opt. Soc. Am. B {\bf 6}, 778 (1989).

\bibitem{zeng} X. C. Zeng, D. J. Bergman, P. M. Hui, and D. Stroud, Phys. Rev.
B {\bf 38}, 10970 (1988).

\bibitem{bruggeman} D. A. G. Bruggeman, Ann. Phys. (Leipzig) {\bf 24}, 636 (1935).

\bibitem{landauer} For reviews, see, e.g., R. Landauer, in
\emph{Electrical Transport and Optical Properties of Inhomogeneous Media}, edited
by J. C. Garland and D. B. Tanner, AIP Conf. Proc. No. 40 (American Institute of
Physics, New York, 1978), pp. 2--45, or D. J. Bergman and D. Stroud, Solid State Phys.
{\bf 46}, 147 (1992).

\bibitem{rammal1} R. Rammal, C. Tannous, P. Breton, and A.-M. S. Tremblay,
Phys. Rev. Lett. {\bf 54}, 1718 (1985).

\bibitem{halperin} B. I. Halperin, Shechao Feng, and P. N. Sen, Phys. Rev. Lett.
{\bf 54}, 2391 (1985).

\bibitem{rammal2} R. Rammal, Phys. Rev. Lett. {\bf 55}, 1428 (1985).

\bibitem{tremblay} A.-M. S. Tremblay, Shechao Feng, and P. Breton, Phys. Rev. B {\bf 33},
2077 (1986).

\bibitem{remillard}  See, e.g., S. K. Remillard, H. R. Yi, and A. Abdelmonem,
IEEE Trans. Appl. Superconductivity {\bf 13}, 3797 (2003).

\end{thebibliography}
\end{document}